\documentclass[aps,prb,twocolumn,preprintnumbers,superscriptaddress]{revtex4}
\usepackage{graphicx}
\usepackage{amsmath}
\usepackage{amssymb}
\usepackage{epstopdf}
\begin{document}

\title{Using neutron spin-echo to investigate proton dynamics in proton-conducting perovskites}
\author{Maths Karlsson}
\affiliation{European Spallation Source Scandinavia,
             Lund University,
             SE-221 00 Lund, Sweden}
\affiliation{Department of Applied Physics,
             Chalmers University of Technology,
             SE-412 96 G{\"o}teborg, Sweden}
\author{Dennis Engberg}
\affiliation{Department of Applied Physics,
             Chalmers University of Technology,
             SE-412 96 G{\"o}teborg, Sweden}
\author{M{\aa}rten E. Bj\"{o}rketun}
\affiliation{Department of Applied Physics,
             Chalmers University of Technology,
             SE-412 96 G{\"o}teborg, Sweden}
\author{Aleksandar~Matic}
\affiliation{Department of Applied Physics,
            Chalmers University of Technology,
            SE-412 96 G{\"o}teborg, Sweden}
\author{G\"{o}ran Wahnstr\"{o}m}
\affiliation{Department of Applied Physics,
             Chalmers University of Technology,
             SE-412 96 G{\"o}teborg, Sweden}
\author{Per G. Sundell}
 \affiliation{Department of Applied Physics,
           Chalmers University of Technology,
            SE-412 96 G{\"o}teborg, Sweden}
\author{Pedro Berastegui}
\affiliation{Department of Inorganic Chemistry, 
             Arrhenius Laboratory, Stockholm University, 
	     SE-106 91 Stockholm, Sweden}
\author{Istaq Ahmed}
\affiliation{Department of Chemical and Biological Engineering, 
             Chalmers University of Technology, 
	     SE-412 96 G{\"o}teborg, Sweden}
\author{Peter Falus}
\affiliation{Institut Laue-Langevin, 
             6 rue Jules Horowitz, BP 156, 38042, 
	     Grenoble Cedex 9, France}
\author{Bela Farago}
\affiliation{Institut Laue-Langevin, 
             6 rue Jules Horowitz, BP 156, 38042, 
	     Grenoble Cedex 9, France}
\author{Lars B{\"o}rjesson}
\affiliation{Department of Applied Physics,
            Chalmers University of Technology,
             SE-412 96 G{\"o}teborg, Sweden}
\author{Sten Eriksson}
\affiliation{Department of Chemical and Biological Engineering, 
             Chalmers University of Technology, 
	     SE-412 96 G{\"o}teborg, Sweden}

\date{\today}


\maketitle
\small

Many hydrated $AB$O$_{3}$-type perovskites are found to be fast proton conductors in the intermediate temperature range $\sim$200--500$^{\circ}$C.  Loading protons in the perovskite structure relies on acceptor-doping to the $B$ site, whereby oxygen vacancies are created, which can be filled with hydroxyl groups in a humid atmosphere at elevated temperatures.\cite{KRE03} On a local scale the proton conduction process is composed of two elementary steps: (i) hydrogen-bond mediated proton transfer between adjacent oxygens, and (ii) rotational motion of the hydroxyl group in between such transfers.\cite{KRE03} The long-range motion of the protons is a series of such transfers and rotations, with an overall rate which depends on the local energy barriers.

The long-range proton diffusion constant is commonly extracted from conductivity experiments via the Nernst-Einstein relation, but there is also one example of the use of pulsed-field gradient nuclear magnetic resonance (PFG-NMR)\cite{KRE96_2}. On microscopic length-scales, proton dynamics in hydrated perovskites have been investigated with frequency-resolved quasielastic neutron scattering (QENS), covering the ps time-scale, extended up to $\sim$1 ns in some cases, in which mainly the local dynamical processes have been observed, although data have also been interpreted in terms of non-localized motions.\cite{HEM95,MAT96,GRO01,KAR09_IN6,KAR96,PIO97,WIL07,BRA08} Investigations of the proton dynamics would, however, benefit greatly if one could extend the time-range to longer time-scales so that the microscopic diffusional process can be studied and information about the influence of the local structure and energy barriers on the proton diffusion can be obtained. This is indeed possible by the use of neutron spin-echo (NSE),\cite{MEZ80,MEZ72,MEZ03} with which the proton dynamics can be studied over a microscopic length-scale in a wide time-range of ps--$\mu$s. However, to our knowledge, this technique has previously neither been applied to study the proton dynamics in hydrated perovskites nor in any other proton-conducting ceramic.

In this Communication we demonstrate the applicability and potential of NSE to study proton dynamics in proton-conducting ceramics. This is exemplified by experiments performed on hydrated BaZr$_{0.90}$Y$_{0.10}$O$_{2.95}$ (10Y:BZO), a cubic perovskite with a relatively high proton conductivity.\cite{KRE03,SCH00} The high proton conductivity together with a high thermodynamic stability make this material a promising candidate for use as electrolyte in intermediate temperature fuel cells.\cite{KRE03} In addition to our experimental work, we use kinetic modeling based on first-principles calculations to assist the interpretation of the experimental data.

The NSE experiment was performed at the IN15 spectrometer at Institut Laue-Langevin (ILL) in Grenoble, France, with which a wide time-range over nearly three decades, $\sim$0.2 to 50 ns, was covered. The measured quantity in NSE is the polarization of the scattered neutrons as a function of the momentum transfer $Q$ and Fourier time $t$, which is directly related to the intermediate scattering function $I(Q,t)$,\cite{MEZ03} revealing the proton dynamics. For instance, for free diffusion the intermediate scattering function would show an exponential relaxation, \textit{i.e.} $I(Q,t)$ = $e$$^{-t/\tau(Q)}$, with the relaxation time $\tau$ proportional to $Q^{2}$, namely $\tau^{-1}$(Q) = $DQ^{2}$ where $D$ is the diffusion coefficient, while a more complex motion would consequently result in a more complex shape of $I(Q,t)$, see further below.\cite{BEE88} Data were obtained for the following temperatures and $Q$-values: 473 K ($Q$ = 0.3 {\AA}$^{-1}$), 521 K ($Q$ = 0.3 {\AA}$^{-1}$), 563 K ($Q$ = 0.15 and 0.3 {\AA}$^{-1}$), 623 K ($Q$ = 0.2 {\AA}$^{-1}$) and 650 K ($Q$ = 0.2 {\AA}$^{-1}$). 

Figure 1 shows the normalized $I(Q,t)$, obtained in the experiment at 521--650 K. As can be seen directly in Figure 
1, $I(Q,t)$ is characterized by a single decay process in the time-range 1--5 ns for all four temperatures. On this time-scale we do not expect any other dynamics than proton motions, since the probed momentum transfers, $Q$ = 0.15--0.3 {\AA}$^{-1}$, corresponds to a length-scale of $\sim$20--40 {\AA} in real space. We can therefore safely assign the relaxation observed in the intermediate scattering function to proton dynamics. Furthermore, we find that the decay in $I(Q,t)$ is well described by a single exponential function for all temperatures, see Figure 1. For the temperature 563 K we have determined the relaxation rate for two $Q$-values and as can be seen in the inset in Figure 1 the relaxation rate follows a $Q^{2}$-behavior. The fact that the relaxation function in the experiment is well described by a single exponential decay, and is consistent with a $Q^{2}$-behavior, shows that we are observing free translational diffusion of the protons.\cite{BEE88} 
This implies that already on a length-scale of 20 {\AA}, which corresponds to a distance of $\sim$5 unit cell lengths (the lattice parameter of 10Y:BZO is 4.20 {\AA})\cite{SCH00}, potential local traps, or other ``imperfections'' in the structure that may severe the proton transport, have averaged out. 
Therefore, we have extracted a diffusion coefficient from the experimentally determined relaxation times assuming a $\tau^{-1}$(Q) = $DQ^{2}$ dependence also for the other temperatures and the result from this analysis is shown in the Arrhenius plot in Figure 2. It is evident that the obtained values are consistent between the different temperatures, which provides support that our analysis is physically reasonable.

\begin{figure} 
\includegraphics[width=0.49\textwidth]{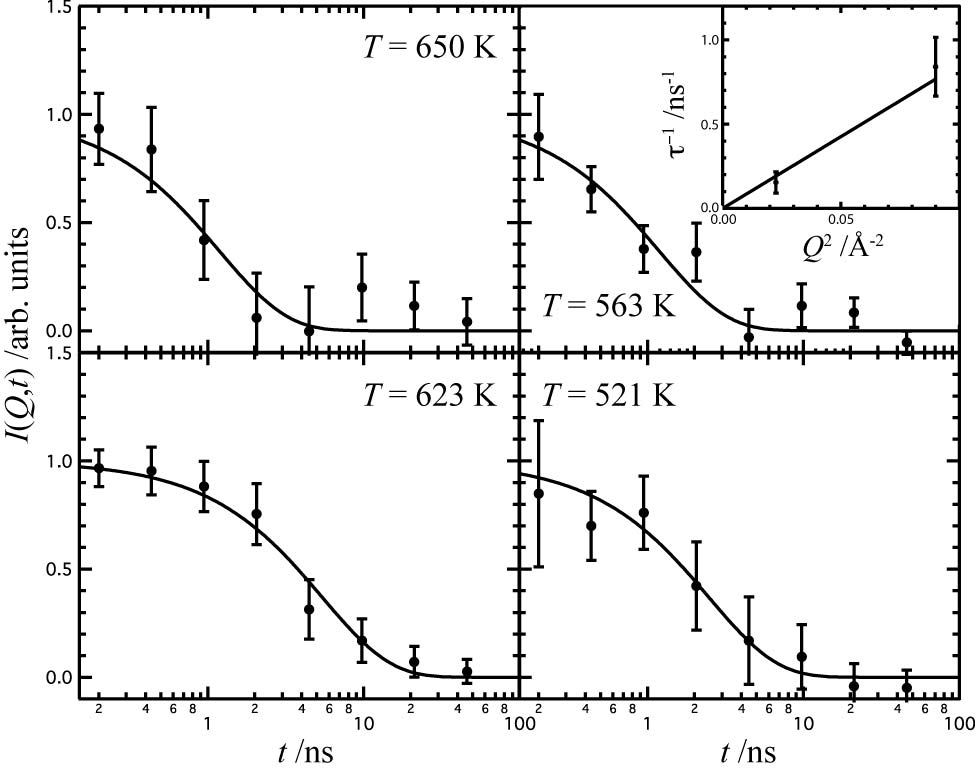}
\caption{\label{Figure1}
$I(Q,t)$ at $T$ = 650 ($Q$ = 0.2 {\AA}$^{-1}$), 623 ($Q$ = 0.2 {\AA}$^{-1}$), 563 ($Q$ = 0.3 {\AA}$^{-1}$) and 521 K ($Q$ = 0.3 {\AA}$^{-1}$). The solid lines are fits to $I(Q,t)$ = $e$$^{-t/\tau(Q)}$. Inset in upper right panel: $Q^{2}$-dependence of the relaxation rate at $T$ = 563 K, where the line represents a fit to $\tau^{-1}$(Q) = $DQ^{2}$ with $D$ = 8.53$\cdot$10$^{-7}$ cm$^{2}$s$^{-1}$.
}
 \end{figure}

To further corroborate our analysis of the NSE data we have also modeled $I(Q,t)$ using a kinetic model based on first-principles calculations.\cite{BJO07,BJO05} The first-principles calculations were carried out within the framework of density functional theory (DFT); details on the settings can be found in ref.\cite{BJO07} and in the Supporting Information. From the DFT calculations the barriers to proton transfer and hydroxyl rotation in the vicinity of and far from dopants were assesed, and used as input parameters for the kinetic model. The calculated diffusion barriers far from Y-dopants are found to be 0.20 eV and 0.18 eV for the proton transfer and hydroxyl rotation motion, respectively. The binding energy to a Y-dopant is 0.16 eV and we find that the influence of the Y-dopant on the energetics for the proton is quite extended in space, including both the first and second coordination shells. For the doping level of 10\% it implies that the region influenced by the dopants amounts to about half of the lattice sites and a conventional description in terms of a trapping diffusion model cannot be used.

Figure 3 shows the calculated intermediate scattering function, $I$$_{\rm{calc.}}$$(Q,t)$, at $T$ = 563 K and for a dopant concentration corresponding to 12.5\%. Results are shown for momentum transfers $Q$ = 0.3, 0.5, 2.0 {\AA}$^{-1}$ and the long-range diffusion limit $Q$ $\rightarrow$ 0, in which the scattering function is given by a single exponential with a characteristic relaxation rate $\tau^{-1}$$(Q)$ = $DQ^{2}$, where $D$ is the calculated diffusion 
coefficient.\cite{BEE88} Since the calculated scattering functions are plotted against $Q^{2}t$ they will collapse onto a single curve as long as we are in the long-range diffusion regime. As seen in Figure 3, this is clearly the case up to, at least, $Q$ = 0.5 {\AA}$^{-1}$. Thus, for the $Q$-values probed in the NSE experiment we indeed expect to observe the long-range diffusional process.

 \begin{figure} 
\includegraphics[width=0.4\textwidth]{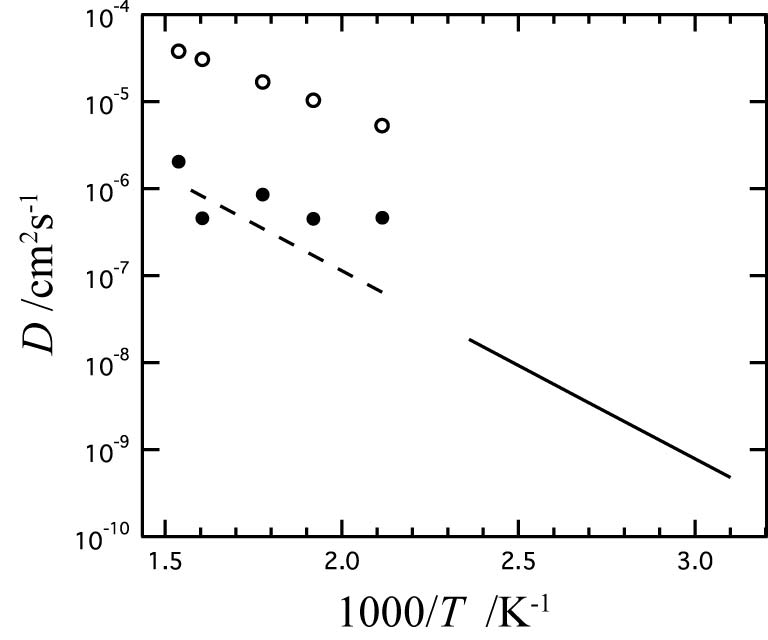}
\caption{\label{Figure2}
Diffusion constant obtained from NSE ($\bullet$), first-principles calcualations ($\circ$), and conductivity measurements (line). The dashed line is an extrapolation of the latter result.
}
 \end{figure}

Only for larger $Q$-values $I_{\rm{calc.}}(Q,t)$ deviates clearly from the single exponential behavior. For example, at $Q$ = 2.0 {\AA}$^{-1}$ in Figure 3 one can observe a two-step relaxation function. This reflects a, so called, trapping behavior, \textit{i.e.} the proton spends an extended time in the vicinity of dopant atoms before it diffuses further through the structure, thus decreasing the long-range proton mobility. The view of dopant atoms acting as well localized trapping centers has previously been proposed based on quasielastic neutron scattering data,\cite{HEM95} muon spin-relaxation experiments,\cite{HEM98} and computer simulations.\cite{DAV99,ISL04,ISL01} It should be noted that for lower dopant concentrations the transition from a single exponential to a more complicated relaxation will occur at smaller $Q$-values.

From the kinetic model based on first-principles calcualtions we have also determined the diffusion constant for five different temperatures. These values are inluded in Figure 2 together with the values obtained from conductivity measurements\cite{KRE01} for comparison. The conductivity measurements were performed at lower temperatures than our NSE measurements, but if one extrapolates the conductivity data we find a good agreement between the proton diffusion constants obtained from the two different experimental techniques. The kinetic model on the other hand overestimates the diffusion constant by about one order of magnitude. However, the accuracy of the obtained potential energy surface is limited by the accuracy of the approximation for the exchange-correlation functional. It is known that generalized gradient approximations (used in the present study) have a tendency to underestimate energy barriers for proton transfer processes.\cite{Barone1996} An underestimation of 0.1 eV, which is not at all unreasonable, would increase the rate by about a factor of 10 at the present temperatures. Changes in the potential energy surface will also influence the value for the vibrational frequencies entering the expression for the prefactors for the  individual jumps. Furthermore, our modeling of the prefactors for the individual rates\cite{BJO07} can be improved by incorporating more vibrational degrees of freedom and the diffusion model we have employed assumes that consecutive proton jumps in the structure are uncorrelated.
 
\begin{figure} 
\includegraphics[width=0.45\textwidth]{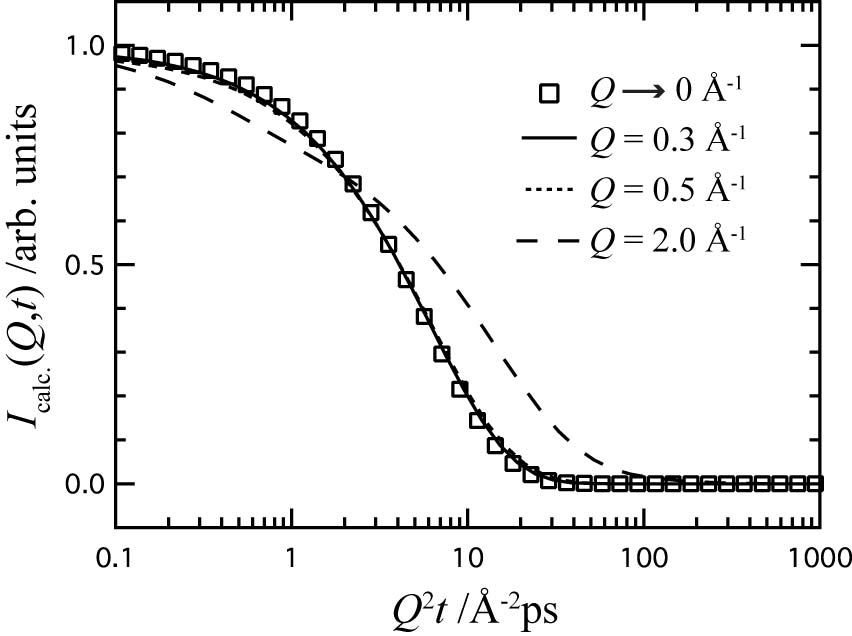}
\caption{\label{Figure3}
Calculated intermediate scattering functions, $I_{\rm{calc.}}$($Q$,$t$), for various momentum transfer values, $Q$, at 563 K.
}
 \end{figure}

The fact that the diffusion constant obtained from the NSE experiment compares well to that derived from the conductivity data\cite{KRE01} is perhaps surprising since the two techniques probe considerably different length-scales. The diffusion constant probed in conductivity measurements reflects the bulk diffusion of protons between grain boundaries, with a length-scale of the order 1 $\mu$m, which is much longer than the length-scale over which the proton diffusion is investigated by NSE, $\sim$20-40 {\AA} in our case. This result suggests that already over a distance as short as five unit cell lengths we are measuring the true long-range proton self-diffusion in the material. Only for $Q$-values larger than 0.5 {\AA}$^{-1}$, corresponding to ~three unit cell lengths, trapping effects are expected in this material. By extending the $Q$-range to higher $Q$-values in the NSE experiment, it should be possible to observe the cross-over from a single exponential at low $Q$, typical for long-range proton diffusion, to a more complex form at larger $Q$-values, suggesting that several processes are reflected in $I(Q,t)$. Such investigations may for example give information about the existence/nature of traps, an often debated subject for hydrated perovskites.
 
To conclude we present data from the first NSE experiment on a proton-conducting perovskite, namely hydrated BaZr$_{0.90}$Y$_{0.10}$O$_{2.95}$, and demonstrate that NSE spectroscopy can be beneficially applied to investigate the protonic self-diffusion in this class of materials, and most likely also in many other types of solid-state proton conductors, such as proton-conducting polymers,\cite{SCH03} solid acids,\cite{HAI01} and alkali thio-hydroxogermanates.\cite{KAR08_Sandals} Furthermore, we show that NSE, which indeed is the only neutron scattering technique that gives access to the long time-scales needed to accurately investigate the comparatively slow proton diffusion in hydrated perovskites, can be powerfully combined with kinetic modeling based on first-principles calculations, since both techniques cover the same time- and length-scales. \\

\textbf{Acknowledgment.} We acknowledge financial support from the Swedish agencies VR, NFSM, and SSF via the ATOMICS program. Allocations of beam time at ILL, and of computer resources through the SNAC are also gratefully acknowledged.\\
 
\textbf{Supporting Information Available:} Details of the sample preparation, the NSE experiment, and the kinetic modeling based on first-principles calculations. This is available free of charge via Internet at http://arXiv.org/.

\end{document}


\title{Using neutron spin-echo to investigate proton dynamics in proton-conducting perovskites - Supporting Information}
\author{Maths Karlsson}
\affiliation{European Spallation Source Scandinavia,
             Lund University,
             SE-221 00 Lund, Sweden}
\affiliation{Department of Applied Physics,
             Chalmers University of Technology,
             SE-412 96 G{\"o}teborg, Sweden}
\author{Dennis Engberg}
\affiliation{Department of Applied Physics,
             Chalmers University of Technology,
             SE-412 96 G{\"o}teborg, Sweden}
\author{M{\aa}rten E. Bj\"{o}rketun}
\affiliation{Department of Applied Physics,
             Chalmers University of Technology,
             SE-412 96 G{\"o}teborg, Sweden}
\author{Aleksandar~Matic}
\affiliation{Department of Applied Physics,
            Chalmers University of Technology,
            SE-412 96 G{\"o}teborg, Sweden}
\author{G\"{o}ran Wahnstr\"{o}m}
\affiliation{Department of Applied Physics,
             Chalmers University of Technology,
             SE-412 96 G{\"o}teborg, Sweden}
\author{Per G. Sundell}
 \affiliation{Department of Applied Physics,
           Chalmers University of Technology,
            SE-412 96 G{\"o}teborg, Sweden}
\author{Pedro Berastegui}
\affiliation{Department of Inorganic Chemistry, 
             Arrhenius Laboratory, Stockholm University, 
	     SE-106 91 Stockholm, Sweden}
\author{Istaq Ahmed}
\affiliation{Department of Chemical and Biological Engineering, 
             Chalmers University of Technology, 
	     SE-412 96 G{\"o}teborg, Sweden}
\author{Peter Falus}
\affiliation{Institut Laue-Langevin, 
             6 rue Jules Horowitz, BP 156, 38042, 
	     Grenoble Cedex 9, France}
\author{Bela Farago}
\affiliation{Institut Laue-Langevin, 
             6 rue Jules Horowitz, BP 156, 38042, 
	     Grenoble Cedex 9, France}
\author{Lars B{\"o}rjesson}
\affiliation{Department of Applied Physics,
            Chalmers University of Technology,
             SE-412 96 G{\"o}teborg, Sweden}
\author{Sten Eriksson}
\affiliation{Department of Chemical and Biological Engineering, 
             Chalmers University of Technology, 
	     SE-412 96 G{\"o}teborg, Sweden}

\date{\today}
\maketitle

\textbf{Sample preparation:}
The BaZr$_{0.90}$Y$_{0.10}$O$_{2.95}$ sample was prepared by mixing stoichiometric amounts of BaCO$_{3}$, ZrO$_{2}$ and Y$_{2}$O$_{3}$. The oxides were heated to 250$^{\circ}$C overnight to remove moisture prior to weighing. Milling was performed manually using an agate mortar and a pestle. The finely ground mixtures were fired at 1000$^{\circ}$C for 8 hours and subsequently ground and pelletized using a 13 mm diameter die under a pressure of 8 tons. The pellets were sintered at 1200$^{\circ}$C in air for 72 hours. After sintering, the pellets were reground, compacted, and refired at 1500$^{\circ}$C for 48 h. Finally, the pellets were finely reground to powders. The charging with protons was performed by annealing the powder samples at ~300$^{\circ}$C under a flow (12 ml/min) of Ar saturated with water vapor at 76$^{\circ}$C for 10 days. X-ray diffraction measurements, performed at ambient temperature on a Siemens D5000 powder diffractometer (CuK$\alpha$$_{1}$ = 1.5406 {\AA}), revealed a cubic structure of the hydrated material, in accordance with the literature [1].\\

\textbf{NSE experiment:}
The NSE experiment was performed at the IN15 spectrometer at Institut Laue-Langevin (ILL) in Grenoble, France. We used neutrons with wavelengths centered at 10 {\AA}. The instrumental resolutions were calibrated using a graphoil elastic scatterer. The data analysis was performed using the standard IN15 data package and FRIDA [2]. The powder sample was loaded in a vacuum tight Al container, which was coated with a 100 nm thick layer of Pt on the inside to avoid corrosion. The sample thickness was chosen to 6 mm, to obtain a total scattering of neutrons of approximately 10\%. It should be noted that the experimental data is customarily normalized through a division with $I$($Q$,$t$ = 0) [3], but in our case this value was close to zero, so normalized scattering functions have instead been obtained from fits of the unnormalized data, $I_{\rm{exp.}}(Q,t)$, according to $I_{\rm{exp.}}(Q,t)=a(Q)+b(Q)\cdot I(Q,t)$, where $a$ and $b$ are $Q$-dependent fitting parameters. For each spectrum, the data was logarithmically summed into 8 data points, in order to increase the statistics.\\

\textbf{Computational details:} 
The first-principles calculations were carried out within the framework of density functional theory, based on the plane-wave pseudopotential approach as implemented in the Vienna \textit{ab-initio} simulation package (VASP) [4,5].  The electron-ion interactions were described by the projector augmented wave method [6]. For the exchange-correlation part, we used a generalized gradient approximation (GGA) due to Wang and Perdew [7]. All calculations were performed non-spin-polarized with a plane-wave cutoff of 400 eV.

In most energy barrier calculations, the hydrated Y-doped BaZrO$_{3}$ sample was represented by a supercell consisting of 3$\times$3$\times$3 primitive BaZrO$_{3}$ units plus one Y ion, which corresponds to a doping level of 3.7\%, and the Brillouin zone sampling was performed using a 2$\times$2$\times$2 k-point grid. In order to ensure the transferability of the results to higher dopant concentrations, similar to that of the ``experimental'' sample, a few key barriers were also calculated using a supercell with 12.5\% doping. Since only minor changes in the barrier heights were observed upon increasing the doping level, we feel confident that reliable predictions about 10Y:BZO can be made based on the 3.7\% data.\\

\noindent
(1) T. Schober, H. G. Bohn, Solid State Ionics 127 (2000) 351.\\ 
(2) http://sourceforge.net/projects/frida (2007). \\
(3) F. Mezei, C. Pappas, T. Gutberlet. Neutron Spin Echo Spectroscopy: Basics, Trends and Applications; Springer: Heidelberg, 2003.\\
(4) G. Kresse, J. Hafner, Phys. Rev. B 48 (1993) 13115.\\
(5) G. Kresse, J. FurthmŸller, Phys. Rev. B 54 (1996) 11169.\\
(6) P. E. Blšchl, Phys. Rev. B 50 (1994) 17953.\\
(7) Y. Wang and J. P. Perdew, Phys. Rev. B 44, 1\\

\bibliography{../../../../../../../../References}